\documentclass[aps,amsmath,amssymb,floatfix,lengthcheck,preprintnumbers,prl,showpacs,superscriptaddress,twocolumn]{revtex4-1}

\usepackage{hyperref}
\usepackage{graphicx}
\usepackage{bm} 
\usepackage{ifpdf} 

\newcommand{\hippy}{\textsc{HiPPy}}
\newcommand{\hpsrc}{\textsc{HPsrc}}
\newcommand{\NR}{\textrm{NR}}
\renewcommand{\vec}[1]{{\bm{#1}}}
\newcommand{\bs}{\mbox{\boldmath $\sigma$}}
\newcommand{\bB}{\mbox{\boldmath $B$}}
\newcommand{\BE}{\begin{equation}}
\newcommand{\EE}{\end{equation}}
\newcommand{\BEA}{\begin{eqnarray}}
\newcommand{\EEA}{\end{eqnarray}}
\newcommand{\BES}{\begin{eqnarray*}}
\newcommand{\EES}{\end{eqnarray*}}
\newcommand{\Psibar}{\overline{\Psi}}
\newcommand{\sla}[1]{\not\!\!#1}

\newcommand{\paperorletter}{letter}

\newcommand{\cambridge}{Department~of~Applied~Mathematics~and~Theoretical~Physics, University~of~Cambridge, Centre~for~Mathematical~Sciences, Cambridge~CB3~0WA, United~Kingdom}
\newcommand{\cray}{Cray~Exascale~Research~Initiative~Europe, JCMB, King's~Buildings, Edinburgh~EH9~3JZ, United~Kingdom}
\newcommand{\edinburgh}{SUPA, School~of~Physics~and~Astronomy, University~of~Edinburgh, King's~Buildings, Edinburgh~EH9~3JZ, United~Kingdom}
\newcommand{\mainz}{Institut~f\"ur~Kernphysik, University~of~Mainz, Becher-Weg~45, 55099~Mainz, Germany}

\begin{document}

\title{Radiative improvement of the lattice NRQCD action using the background field method and application to the hyperfine splitting of quarkonium states}

\author{T.~C. \surname{Hammant}}
\affiliation{\cambridge}
\author{A.~G. \surname{Hart}}
\affiliation{\cray}\affiliation{\edinburgh}
\author{G.~M. \surname{von Hippel}}
\affiliation{\mainz}
\author{R.~R. \surname{Horgan}}
\author{C.~J. \surname{Monahan}}
\affiliation{\cambridge}

\preprint{DAMTP-2011-35}
\preprint{MKPH-T-11-11}

\begin{abstract} 
We present the first application of the background field method
to Non-Relativistic QCD (NRQCD) on the lattice
in order to determine the one-loop radiative corrections
to the coefficients of the NRQCD action
in a manifestly gauge-covariant manner.
The coefficient of the $\vec{\sigma}\cdot\vec{B}$ term in the NRQCD action
is computed at the one-loop level;
the resulting shift of the hyperfine splitting of bottomonium
is found to bring the lattice predictions in line with experiment.
\end{abstract}

\maketitle


Non-Relativistic QCD~(NRQCD)
\cite{Lepage:1992}
is an effective field theory that has been applied with considerable success
to the description of hadrons containing heavy quarks
\cite{Gray:2005ur}.
However, the currently used NRQCD actions do not include radiative improvement
(with the exception of tadpole improvement), and this will affect the precision
with which crucial quantities such as the hyperfine splitting between the
$\Upsilon$ and the $\eta_{\rm b}$ can be determined
\cite{Meinel:2010pv}.
In contrast to this, Non-Relativistic QED~(NRQED) has been successfully improved
and applied to obtain highly precise theoretical predictions for the fine
structure of muonium
\cite{Kinoshita:1995mt,Nio:1997fg}.
It is clearly highly desirable to improve NRQCD in a similar manner. This is
rendered complicated by the nature of the non-abelian gauge interactions in
QCD and NRQCD, which requires NRQCD to be implemented on a lattice and hence
makes it necessary to retain the full $1/(ma)^n$ dependences, whereas in
continuum NRQED $(\Lambda/m)^n$ terms can be omitted in a consistent manner.
Moreover, IR divergences play a non-trivial r\^ole in QCD and NRQCD.

In this \paperorletter, we present the first calculation of radiative
corrections to coefficients in the lattice NRQCD action using
the background field~(BF) method. We compute the one-loop effective action in lattice
NRQCD and match it term-by-term to the non-relativistic reduction of the one-loop effective
action for continuum QCD. In particular, we determine the one-loop corrections to the 
coefficient of the chromomagnetic $\vec{\sigma}\cdot\vec{B}$ term  and the four-fermion
spin-spin interaction; these corrections are important for the accurate
calculation of the hyperfine structure of heavy quark states using NRQCD.

\section{The Background field method for lattice NRQCD}
\label{sec:BFNRQCD}

The BF method
\cite{Abbott:1983,DeWitt:1967ub,DeWitt:1967uc,KlubergStern:1974xv}
is a well-established tool to compute the effective action in quantum
field theory. The auxiliary gauge invariance of BFG amplitudes implies that 
the effective action contains only gauge-covariant operators which leads to a 
set of Ward Identities in QCD that reduce the amount of calculation necessary to 
renormalize the theory.  This property is important for operators of dimension 
$D>4$ where the loss of gauge-covariance would lead to a proliferation of 
additional operators  and is vital to the radiative improvement of NRQCD which is
a non-relativistic expansion on operators of increasing dimension; only BFG will
guarantee the gauge covariance of the improved effective action.
Whilst the presence of gauge non-covariant finite terms with $D>4$ in the effective action is 
not {\it per se} incorrect, they obscure the underlying gauge symmetry and greatly complicate 
the theory and simulation. An attempt to match without using BFG would lead to the appearance
of ultraviolet logarithms, which would have to be cancelled by the contributions
from additional non-gauge-covariant operators. Although BFG does not guarantee that the
coefficients in the effective action are independent of the gauge parameter
\cite{Rebhan:1986wp},
in our case we match between theories using on-shell quantities and we explicitly find 
that the coefficients are independent of the gauge parameter in both QCD and NRQCD.
Moreover, the QED-like Ward identities in BFG imply that the one-particle
irreducible~(1PI) vertex functions are finite, and that the coupling $g$ is
renormalized only by the contribution from the gluonic self-energy, 
whereas the BF is not renormalized. This is true both in QCD and NRQCD, which allows us 
to match the theories by equating two finite quantities. As a consequence of this crucial property
of BFG, we are free to use different regulators in QCD and NRQCD. In particular,
we can calculate the QCD vertex analytically in the continuum using dimensional
regularization, or on a fine lattice and taking the continuum limit. The latter
is particularly convenient for checking the gauge-parameter independence of the
result, since the analytical calculation becomes rather involved for general
values of the gauge parameter.

In the following we denote the perturbative expansion for a generic parameter $w$
as $w = \sum_{n=0}w^{(n)}\alpha^n$.

\section{\texorpdfstring{Matching the $\bs\cdot\bB$ term}{Matching the s.B term}}
\label{sec:sigmadotB}

\begin{figure}
\includegraphics[width=0.83\columnwidth,keepaspectratio=]{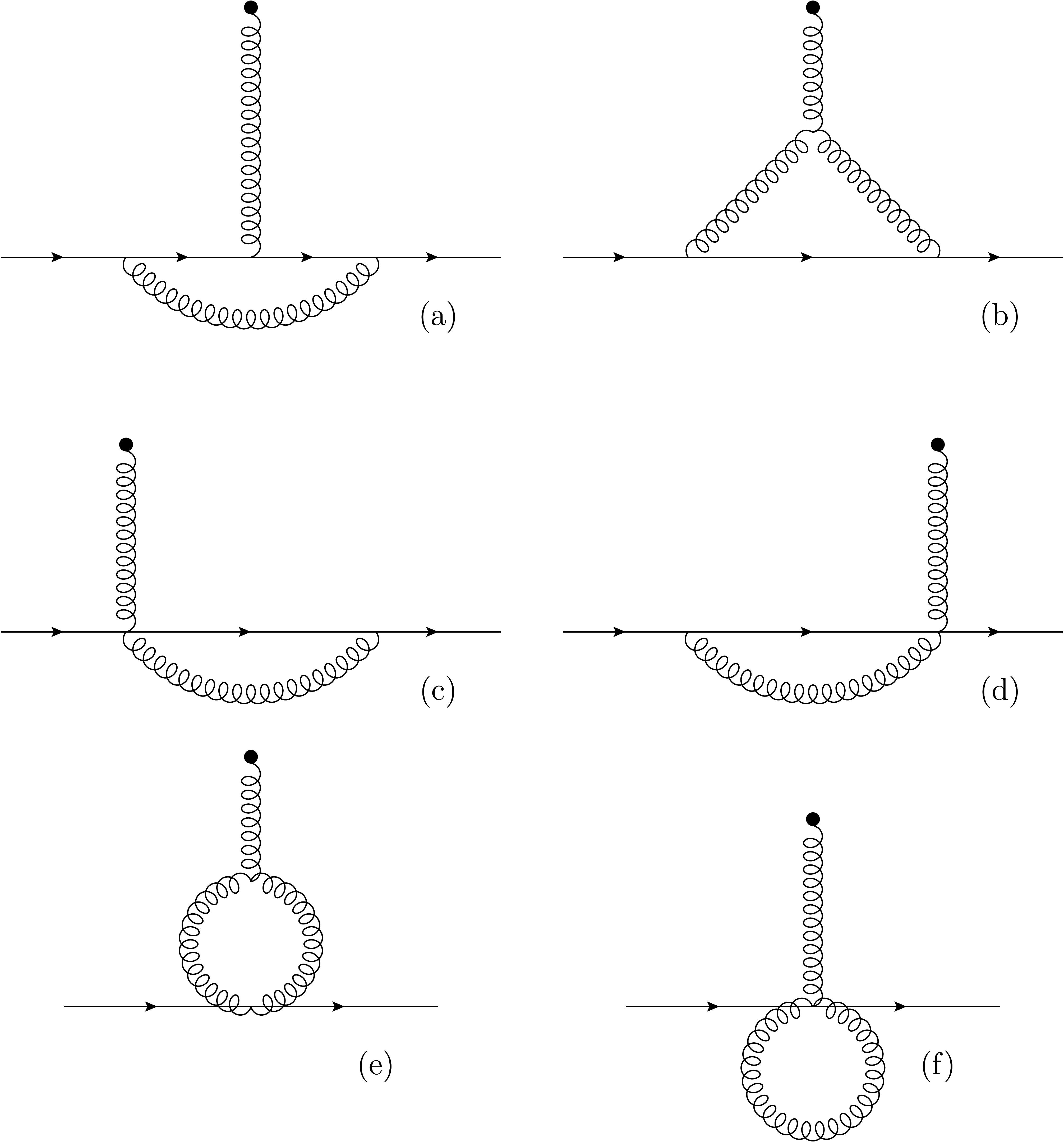}
\caption{Feynman diagrams to be computed in both QCD and NRQCD for matching the 
$\vec{\sigma}\cdot\vec{B}$ term in the NRQCD action.}
\label{fig:BFdiagrams}
\end{figure}

The effective action for continuum QCD contains the
following terms involving the fermion fields:
\BE
\Gamma[\Psi,\Psibar,A] = Z_2^{-1}\Psibar\sla{D}\Psi + \delta Z_\sigma \Psibar
\frac{\sigma^{\mu\nu}F_{\mu\nu}}{2m}\Psi + \ldots
\EE
which after renormalization of the first term gives
\BE
\Gamma[\Psi_R,\Psibar_R,A] = \Psibar_R\sla{D}\Psi_R + b_\sigma\Psibar_R
\frac{\sigma^{\mu\nu}F_{\mu\nu}}{2m_R}\Psi_R + \ldots
\EE
with
\BE
b_\sigma = \delta Z_\sigma Z_2 Z_m = \sum_{n=1} b_\sigma^{(n)} \alpha^n\;,
\EE
where the leading correction is of order $O(\alpha_s)$ and comes from
$\delta Z_\sigma$ alone. After performing the
non-relativistic reduction by a Foldy-Wouthuysen-Tani (FWT) transformation,
we find that the term relevant for the determination of the chromomagnetic
moment of the quark is
\BE
\left(1+b_\sigma\right)\psi_R^\dag\frac{\vec{\sigma}\cdot\vec{B}}{2m_R}\psi_R\;.
\EE
A straightforward analytical calculation of the Feynman diagrams shown in
figure
\ref{fig:BFdiagrams} (a)--(b)
gives
\BE
b_\sigma = \left(\frac{3}{2\pi}\log\frac{\mu}{m}+\frac{13}{6\pi}\right)\alpha
\EE
at the one-loop level, where $\mu$ is the infrared cutoff.

The effective action for NRQCD contains the spin-dependent term
\BE
\Gamma_\sigma[\psi,\psi^\dag,A] = c_4Z_\sigma^{\NR}\psi^\dag\frac{\vec{\sigma}\cdot\vec{B}}{2M}\psi
\EE
which after renormalization becomes
\BE
\Gamma_\sigma[\psi_R,\psi_R^\dag,A] = 
  c_4Z_\sigma^{\NR}Z_2^{\NR}Z_m^{\NR}\psi_R^\dag\frac{\vec{\sigma}\cdot\vec{B}}{2M_R}\psi_R\;.
\EE
We require that the anomalous chromomagnetic moment in QCD and NRQCD be equal and find the matching condition
\BE
c_4 Z_\sigma^{\NR} Z_2^{\NR} Z_m^{\NR} = 1 + b_\sigma
\EE
and at tree level and one-loop order we find
\BEA
c_4^{(0)} &=& 1\;, \\ \nonumber
c_4^{(1)} &=& b_\sigma^{(1)}-\delta Z_\sigma^{\NR,(1)}-\delta Z_2^{\NR,(1)}-\delta Z_m^{\NR,(1)}\;.
\EEA
The NRQCD contribution to $c_4^{(1)}$ contains a logarithmic IR divergence $\frac{3\alpha}{2\pi}\log(\mu a)$, 
which combines with the IR logarithm from the QCD result above to yield an overall logarithmic 
contribution $-\frac{3\alpha}{2\pi}\log(M a)$.

Besides the ordinary diagrammatic contributions calculated below, we also need to take 
into account the contributions from the mean-field improvement
$U\mapsto U/u_0$, which affect $\delta Z_\sigma^{\NR,(1)}$ and
$\delta Z_m^{\NR,(1)}$. Perturbatively, $u_0=1-\alpha_s u_0^{(2)}$, and the
contributions from inserting this expansion into the NRQCD action
can be worked out algebraically. The final result for the one-loop correction to $c_4$ is then
\BEA
c_4^{(1)} &=&\textstyle \frac{13}{6\pi}-\delta \tilde Z_\sigma^{\NR,(1)}-\delta \tilde Z_2^{\NR,(1)}-
\delta \tilde Z_m^{\NR,(1)}\;, \\ \nonumber 
&-&\textstyle \delta Z_m^{\rm tad,(1)} - \delta Z_\sigma^{\rm tad, (1)} -\frac{3}{2\pi} \log Ma 
\label{eq:c4}
\EEA
where $\delta \tilde Z_X$ denotes a finite diagrammatic contribution.
We expect the coefficient $c_4$ to be gauge-parameter independent for on-shell quarks, since it is directly 
related to the hyperfine splitting, which is a physical quantity.

\section{The four-fermion spin-spin interaction}
\label{sec:four_fermion}
In NRQCD the hyperfine splitting in the $b\bar{b}$ system also receives a contribution
from the spin-dependent four-fermion operators generated by $Q\bar{Q} \to Q\bar{Q}$ scattering 
in the colour singlet channel. It is conventional to write these contributions using a 
Fierz transformation \cite{Lepage:1992,Labelle:1997}
\BE
S_{4f} = d_1\frac{\alpha^2}{M^2}(\psi^\dagger \chi^*)(\chi^T \psi) 
 + d_2\frac{\alpha^2}{M^2}(\psi^\dagger \vec{\sigma}\chi^*)\cdot (\chi^T \vec{\sigma}\psi)\;,
\label{eqn:4f}
\EE
where $\psi$ and $\chi$ are the quark and anti-quark fields, respectively, treated as
different particle species with corresponding representations of their spin and colour algebras. 
The spin-independent contributions to $d_1$ and $d_2$ from $Q\bar{Q}$ scattering are not
included as they do not influence the hyperfine structure. In QCD the two continuum diagrams are
shown in figures \ref{fig:4fdiagrams}(a) and \ref{fig:4fdiagrams}(b), and in NRQCD all diagrams in 
figure \ref{fig:4fdiagrams} need to be calculated. The one-loop contributions to the renormalization 
constants for the operators in eqn. (\ref{eqn:4f}) take the form, respectively,
\BEA
Z_{f1}&=&\textstyle\alpha^2\left(A_{f1} - \log \frac{\mu}{m} - 
\frac{16\pi}{27}\frac{m}{\mu}\right)\;, \nonumber \\
Z_{f2}&=&-\frac{1}{3}Z_{f1}.
\EEA
The last term in both expressions is the Coulomb singularity arising from the Coulomb gluon exchange
in figure \ref{fig:4fdiagrams}(a). For QCD these expressions were verified numerically and for both QCD
and NRQCD were shown to be gauge-parameter independent; there are two independent colour trace
combinations, each of which is separately gauge independent. In the numerical calculations we used
IR subtraction functions to analytically remove both IR and Coulomb divergences; this greatly improved convergence. 
For QCD we find 
\BE
A^R_{f1}~=~{\frac{8}{9}}\;.
\EE
The matching parameters for the term in the NRQCD action, including the two-gluon 
annihilation contribution to $d_1$ \cite{Labelle:1997}, are then
\BEA
d_1&=&  -3d_2 - \frac{2}{9}(2-2\log 2 )\;, \nonumber \\
d_2&=&  -\frac{8}{27} + \frac{1}{3}A^{NR}_{f1} - \frac{1}{3}\log{Ma}\;.
\label{eq:d}
\EEA

\section{Implementation and results}
\label{sec:results}

To perform the calculation in NRQCD, we employ the \hippy\ and \hpsrc\ packages
for automated lattice perturbation theory
\cite{Hart:2004bd,Hart:2009nr},
which we extended to deal with the modifications of the usual Feynman rules
engendered by the use of BFG
\cite{Luscher:1995vs,Luscher:1995np}.
Specifically, in the expansion of a gauge link the background fields $B_\mu$
must always appear to the right of the quantum fields $A_\mu$ and so not all orderings
of background and quantum fields can arise. Additional contributions to all 
purely gluonic vertices including exactly two quantum gluons arise from the gauge-fixing term
and additional ghost field vertices are generated which have
been included but are not needed for the present calculation.
For further implementation details the reader is referred to
\cite{Hammant:2010aq}.

\begin{figure}
\includegraphics[width=0.83\columnwidth,keepaspectratio=]{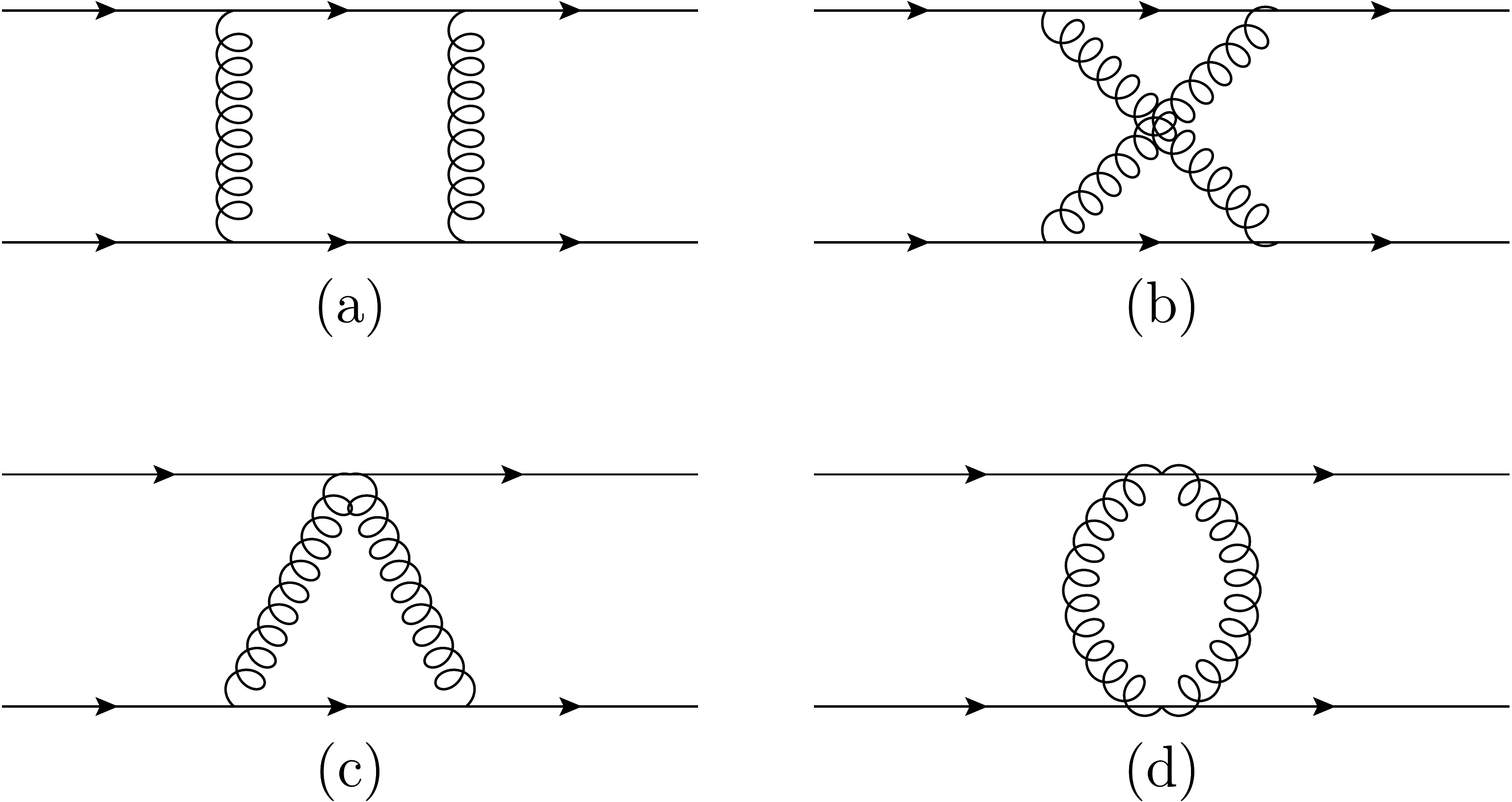}
\caption{Feynman diagrams to be computed in both QCD and NRQCD for matching the
four-fermion terms in the NRQCD action. There are two diagrams with the topology of (c).}
\label{fig:4fdiagrams}
\end{figure}

For the $\vec{\sigma}\cdot\vec{B}$ operator matching we compute the diagrams in figures
\ref{fig:BFdiagrams} (a)--(f)
and for the four-fermion operator matching we compute the diagrams in figure
\ref{fig:4fdiagrams}.
We use the \hpsrc\ library, which includes a parallel implementation
of VEGAS
\cite{Lepage:1977sw},
as well as routines for automatic differentiation of Feynman diagrams
\cite{vonHippel:2009zz}.
We carried out a number of checks of the calculation.
Firstly, we replicate the known IR logs correctly. We find
that the coefficients of these logs are gauge-parameter independent and, since
this is not true of the contributions from individual diagrams, it provides
a strong check. Second, we check that the non-logarithmic part
of the result is similarly gauge-parameter independent where the
individual contributions are not. For matching the four-fermion terms it is vital to 
employ IR subtraction functions to remove logarithmic and Coulomb IR singularities.

\begin{table}
\begin{tabular}{|c|ccccc|}
\hline
$Ma$ & 1.95 &~~& 2.8 &~~& 4.0 \\\hline\hline
$\delta\tilde Z_\sigma+\delta\tilde Z_2$ & -5.164(7) && -4.913(6) && -4.739(6)
\\
$\delta\tilde Z_m$ & 1.512(1)  && 1.022(3) && 0.723(2) \\
$\delta Z_\sigma^{\rm tad}$ & 4.387 && 4.077 && 3.841 \\
$\delta Z_m^{\rm tad}$ & -1.092 && -0.787 && -0.641  \\
$c_4^{(1)}$ & 0.728(7) && 0.799(7) && 0.842(6) \\
$d_1$ & 0.638(7) && -0.109(14) && -1.138(25) \\
$d_2$ & -0.258(2) && -0.009(5) && 0.334(8) \\\hline
\end{tabular}
\caption{Renormalization parameters of the $\vec{\sigma}\cdot\vec{B}$
and the four-fermion terms defined, respectively, in eqns. (\ref{eq:c4}) and
(\ref{eq:d}).}
\label{tab:results}
\end{table}

For NRQCD, we used the action from
\cite{Gray:2005ur}
with stability parameter $n=2$, and we used the Symanzik improved gluon action
\cite{Luscher:1986},
which were also used by the MILC collaboration \cite{MILC:2001} whose configurations were used in \cite{Gray:2005ur}. 
We find
\BE\textstyle
\delta Z_m^{\rm tad,(1)} = -\left(\frac{2}{3} +\frac{3}{(Ma)^2}\right)\alpha_s u_0^{(2)}
\EE
The tadpole contribution to $\delta Z_\sigma^{\NR,(1)}$ comes from the
mean-field improvement of the improved field-strength tensor and from
the cross-multiplication of the tree-level $\vec{\sigma}\cdot\vec{B}$ term with the tadpole corrections
terms in $H_0$
\cite{Gray:2005ur}.
The overall result is
\BE\textstyle
\delta Z_\sigma^{\rm tad, (1)} = \left(\frac{13}{3}+\frac{13}{4Ma}-\frac{3}{8 n (Ma)^2}-\frac{3}{4(Ma)^3}\right)\;.
u_0^{(2)}\;.
\EE

We chose the Landau mean link to be $u_0^{(2)} = 0.750$ \cite{Nobes:2001}.
Our results are shown in table
\ref{tab:results}.
\begin{table}
\begin{tabular}{|cc|cc|c|c|}
\hline
&&\multicolumn{2}{|c|}{~Correction \%~}&hfs (MeV)&hfs (MeV)\\
$Ma$  & $\alpha_V(q^*)$ & 4-fermion&$\sigma\cdot B$& ref.
\cite{Gray:2005ur}&corrected\\\hline\hline
1.95  & 0.216        & -10.3(1) & +31.4(3)  &56(2)&68(3)(5)(6) \\
2.8   & 0.249        & +1.3(2) &  +39.8(3) &50(2)&71(3)(6)(5) \\
4.0   & 0.293        & +23.2(4) &  +49.3(3) &41(2)&71(3)(7)(4) \\\hline
\end{tabular}
\caption{Corrections to the bottomonium hyperfine splitting results
of \cite{Gray:2005ur} arising from the radiative improvement of the action. In
the
last column the errors are statistical, $O(\alpha^2)$, and relativistic
corrections.}
\label{tab:hyperfine}
\end{table}
Whilst there is no substitute for including these radiative corrections in
a simulation, we note that both operators give a contribution to the hyperfine splitting
that is dominated by a contact term. The leading contribution from the $g\bs\cdot\bB$ term,
already included in the simulation, is $O(\alpha_s)$ and so the radiative correction to $c_4$
and the leading contribution from the four-fermion terms in eqn. (\ref{eqn:4f}) both contribute at
$O(\alpha^2)$ giving an $O(\alpha)$ correction to the measured tree-level contribution. A reasonable 
estimate for the multiplicative correction to the tree-level prediction for the hyperfine splitting 
is then 
\BE
1~+~\alpha_V(q^*)\left(2\,c_4^{(1)}-\frac{27}{16\pi}\left(d_1-d_2\right)\right)\;,
\EE
where we chose $q^*=\pi/a$. Applying our results to the hyperfine splitting of bottomonium, we find
the corrections given in table
\ref{tab:hyperfine}
for the data points of
\cite{Gray:2005ur}.
On all lattices the correction is positive and the remaining $O(a^2)$ error in 
the NRQCD predictions of \cite{Gray:2005ur} is reduced to be within errors.

\section{Conclusion}
\label{sec:conclusions}

In this \paperorletter, we have applied the BF method to lattice NRQCD for the first time and have computed 
the one-loop radiative correction to the coefficient, $c_4$, of the $\vec{\sigma}\cdot\vec{B}$ operator and 
the one-loop radiative contribution to the coefficients, $d_1$ and $d_2$ of the four-fermion contact
operators that affect the hyperfine structure of heavy quark mesons. The gauge independence of our
calculation was explicitly checked by carrying out both relativistic and non-relativistic calculations
in the lattice theory. This is possible because in BFG all calculations are UV finite. Our results are summarized in 
table \ref{tab:results} and in eqns. (\ref{eq:c4}) and (\ref{eq:d}). In particular, in eqn. (\ref{eq:c4}) there is a 
negative correction to $c_4$ due the IR divergences. However, it turns out that the constant terms 
more than cancel this effect and the correction to $c_4$ is positive. Whilst there is no substitute for 
including these corrections in a simulation, we have given an estimate for the correction to the 
$\Upsilon - \eta_b$ hyperfine splitting measured by Gray et al. \cite{Gray:2005ur} in table \ref{tab:hyperfine}.  
The result is to reduce the lattice spacing dependence to within errors and to give an estimate for this
hyperfine splitting of $68(3)(5)(6)$MeV to be compared with the experimental value of $69.3(2.8)$MeV \cite{PDG:2010}.
The errors shown are statistical, $O(\alpha^2)$, and due to relativistic corrections, respectively.
The elimination of $O(\alpha a^2)$ errors and the agreement with experiment gives us confidence that the
calculations are robust.

The determination of the one-loop radiative corrections to the coefficients of the $\vec{p}^4$, Darwin and 
spin-orbit terms and other four-fermion contact terms, as well as more details of the calculations 
will be presented in a longer paper in the near future.

\acknowledgments
We thank Alan Gray, Andrew Lee, Christine Davies and Matthew Wingate
for useful discussions.
We thank the DEISA Consortium,
co-funded through the EU~FP6 project RI-031513 and the FP7 project RI-222919,
for support within the DEISA Extreme Computing Initiative. 
This work was supported by STFC under grants ST/G000581/1 and ST/H008861/1.
The calculations for this work were, in part, performed on the University
of Cambridge HPCs as a component of the DiRAC facility jointly funded by
STFC and the Large Facilities Capital Fund of BIS.
The University of Edinburgh is supported in part by the 
Scottish Universities Physics Alliance~(SUPA).
AH was supported in part by the Royal Society~(UK).

\bibliographystyle{h-physrev4}
\bibliography{sigma_B}

\end{document}